\documentclass[iop]{emulateapj}

\usepackage{upgreek}
\usepackage{amssymb,amsmath}
\usepackage{graphicx}
\usepackage{multirow}
\usepackage{natbib}
\usepackage{threeparttable}
\def\citeN{\citet}
\def\cite{\citep}

\def\apj{ApJ}                 % Astrophysical Journal
\def\aap{A\&A}                % Astronomy and Astrophysics
\def\mnras{MNRAS}             % Monthly Notices of the RAS
\def\nat{Nature}              % Nature
\usepackage[usenames,dvipsnames]{color}

\shorttitle{An Absence of Fast Radio Bursts at Intermediate Galactic Latitudes}
\shortauthors{Petroff et al.}

\begin{document}

\title{An Absence of Fast Radio Bursts at Intermediate Galactic Latitudes}

\author{E.~Petroff\altaffilmark{1,2,3}, W.~van Straten\altaffilmark{1,3}, S.~Johnston\altaffilmark{2}, M.~Bailes\altaffilmark{1,3}, E.~D.~Barr\altaffilmark{1,3}, S.~D.~Bates\altaffilmark{4}, N.~D.~R.~Bhat\altaffilmark{3,5}, M.~Burgay\altaffilmark{6}, S.~Burke-Spolaor\altaffilmark{7}, D.~Champion\altaffilmark{8}, P.~Coster\altaffilmark{1}, C.~Flynn\altaffilmark{1}, E.~F.~Keane\altaffilmark{1,3}, M.~J.~Keith\altaffilmark{4}, M.~Kramer\altaffilmark{8,4}, L.~Levin\altaffilmark{9}, C.~Ng\altaffilmark{8}, A.~Possenti\altaffilmark{6}, B.~W.~Stappers\altaffilmark{4}, C.~Tiburzi\altaffilmark{6,10}, D.~Thornton\altaffilmark{4}}

\email{epetroff@astro.swin.edu.au}

\altaffiltext{1}{Centre for Astrophysics and Supercomputing, Swinburne University of Technology, P.O. Box 218, Hawthorn, VIC 3122, Australia}
\altaffiltext{2}{CSIRO Astronomy \& Space Science, Australia Telescope National Facility, P.O. Box 76, Epping, NSW 1710, Australia}
\altaffiltext{3}{ARC Centre of Excellence for All-sky Astrophysics}
\altaffiltext{4}{Jodrell Bank Centre for Astrophysics, University of Manchester, Alan Turing Building, Oxford Road, Manchester M13 9PL, United Kingdom}
\altaffiltext{5}{ICRAR/Curtin University, Curtin Institute of Radio Astronomy, Perth WA 6845, Australia}
\altaffiltext{6}{INAF - Osservatorio Astronomico di Cagliari, Via della Scienza, 09047 Selargius (CA), Italy}
\altaffiltext{7}{Jet Propulsion Laboratory, California Institute of Technology, 4800 Oak
Grove Drive, Pasadena CA 91104 USA}
\altaffiltext{8}{Max Planck Institut f\"{u}r Radioastronomie, Auf dem H\"{u}gel 69, 53121 Bonn, Germany}
\altaffiltext{9}{Department of Physics, West Virginia University, Morgantown, WV 26506, USA}
\altaffiltext{10}{Dipartimento di Fisica, Universit\`a di Cagliari, Cittadella Universitaria 09042 Monserrato (CA), Italy}

\begin{abstract}
Fast radio bursts (FRBs) are an emerging class of bright, highly dispersed radio pulses. Recent work by Thornton et al. (2013) has revealed a population of FRBs in the High Time Resolution Universe (HTRU) survey at high Galactic latitudes. A variety of progenitors have been proposed including cataclysmic events at cosmological distances, Galactic flare stars, and terrestrial radio frequency interference. Here we report on a search for FRBs at intermediate Galactic latitudes ($-15^{\circ}$ $< b <$ 15$^{\circ}$) in data taken as part of the HTRU survey. No FRBs were discovered in this region. Several effects such as dispersion, scattering, sky temperature and scintillation decrease the sensitivity by more than 3$\sigma$ in $\sim$20\% of survey pointings. Including all of these effects, we exclude the hypothesis that FRBs are uniformly distributed on the sky with 99\% confidence. This low probability implies that additional factors -- not accounted for by standard Galactic models -- must be included to ease the discrepancy between the detection rates at high and low Galactic latitudes. A revised rate estimate or another strong and heretofore unknown selection effect in Galactic latitude would provide closer agreement between the surveys' detection rates. The dearth of detections at low Galactic latitude disfavors a Galactic origin for these bursts. 
\end{abstract}


\section{Introduction}\label{sec:intro}

Fast radio bursts (FRBs) are single, bright, highly dispersed radio pulses of millisecond duration. These bursts have fluences of 0.6 - 8.0 Jy ms, and dispersion measures (DMs) well in excess of the expected Galactic contribution along the line of sight. Given the high flux density and high DM-derived distances, they could arise from high luminosity events at $z \lesssim$ 1.

The first burst of extragalactic origin \cite{Lorimer07} was discovered using the Parkes multibeam receiver \cite{multibeam} in a pulsar survey of the Magellanic Clouds at Galactic latitude $b = -41^{\circ}$. Subsequently, the astrophysical origin of the Lorimer burst was called into question by the discovery of apparently dispersed sources in archival surveys called ``perytons'', which are still poorly understood \cite{Burke11,Bagchi12}. Perytons exhibit dispersive properties that are similar, though not identical, to those of an astrophysical source; however, they are detected in all 13 beams of the Parkes 21-cm multibeam receiver, which is impossible for a distant point source. They have peculiar frequency structure, with all perytons showing bright `nodules' in specific regions of the observing band. Perytons may be an atmospheric phenomenon or terrestrial radio-frequency interference (RFI) of unknown origin \cite{Kocz12,KatzPerytons}.

A recent analysis of high latitude observations from the High Time Resolution Universe survey (HTRU) by \citeN{Thornton13} revealed a population of four new FRB sources. Currently there are seven published FRB events ranging in DM from 375 to 1106.7 pc cm$^{-3}$ \cite{Lorimer07,Keane12,Thornton13,Spitler14}. Most FRBs have been found at Galactic coordinates where the Galactic contribution to DM is small ($<$ 100 cm$^{-3}$ pc).

One of the scenarios considered in recent work by \citet{Shri} is a terrestrial origin for FRBs, where they are interpreted as being similar to peryton events, but occuring at greater distances ($>$ 40km) from the telescope, mimicking a source at infinity. In this model the Lorimer burst, which was detected in three adjacent beams of the multibeam receiver, occupies a place between traditional FRB events and traditional peryton events and is believed to have occured at some distance from the detector close to the Fresnel scale for Parkes, 20 km. 

\citet{Loeb2014} have proposed that FRBs originate in the envelope of main-sequence flare stars within our Galaxy, after finding a flare star within the full width at half maximum of the detection beam of one FRB from the \citet{Thornton13} sample. Recently, however, several authors have highlighted why such a model is physically untenable \cite{Luan,Dennison14}.

Several progenitors at cosmological distances have also been proposed, including soft gamma-ray repeaters \cite{Thornton13,Shri}, and the collapse of supramassive neutron stars \cite{Falcke} located in distant host galaxies. Each of these mechanisms is theorized to be capable of producing coherent radio emission of millisecond duration.

At present, archival and on-going radio pulsar surveys are the most immediate and obvious places to begin searching for more FRBs. \citeN{Thornton13} estimate an FRB rate of R$_\mathrm{FRB}$ $\sim$ 1.0$\substack{+0.6 \\ -0.5} \times 10^{4}$ sky$^{-1}$ day$^{-1}$ for bursts with fluences $F \sim 3$ Jy ms, based on 615 hours of observations. The HTRU intermediate latitude survey used an identical setup to \citet{Thornton13} with 1,157 hours on-sky.

Here we report on a search for FRBs in the HTRU intermediate latitude survey in a DM range from 100 - 5000 cm$^{-3}$ pc which returned no new, highly-dispersed pulses. An introduction to the HTRU survey, the analysis tools, and results of the single pulse search are presented in Section~\ref{sec:analysis}. We present an in-depth discussion of our non-detection in Section~\ref{sec:discussion}.

\section{Analysis and Results}\label{sec:analysis}
The High Time Resolution Universe (HTRU) Survey was designed as a comprehensive survey of the radio sky with 64-$\upmu$s time resolution at 1.4~GHz to detect pulsars and other transient radio phenomena. Observations for HTRU are divided into three observing regimes at low, intermediate, and high Galactic latitudes. This study focuses on the intermediate latitude component of the survey: 540-s pointings in the range $-120^{\circ} < \ell < 30^{\circ}$ and $|b| < 15^{\circ}$ with the 13-beam Parkes 21-cm multibeam receiver, which has a 0.5 deg$^{2}$ field of view.

The survey data were searched for single pulses using techniques similar to those described in \citet{BurkeSpolaorSP}. The dedispersion and width filter matching were optimized for processing on a graphics processing unit (GPU) with the new software package \textsc{Heimdall}\footnote{http://sourceforge.net/projects/heimdall-astro/}. \textsc{Heimdall} produces a list of candidates for each beam.

All 13 beams in a single pointing are then run through a coincidence detector. Candidates that occur in more than 3 beams, fewer than 3 adjacent DM trials, have a S/N $<$ 6 or DM $<$ 1.5 cm$^{-3}$ pc are rejected. With these cuts we are not sensitive to peryton-like events that occur in all 13 beams of the receiver. We will address the topic of perytons in the HTRU survey in a subsequent paper. The candidates that remain after these cuts are concatenanted into a single candidate list for the entire pointing. Each pointing candidate file was searched for FRBs by looking for candidates matching the following criteria

% \begin{equation}\label{eq:thresholds}
% \begin{split}
% &\mathrm{S/N} > 10 \\
% & \mathrm{W} \leq 2^8 \times 64 \mathrm~\upmu \mathrm{s} = 16.3 ~\mathrm{ms} \\
% &\mathrm{DM ~ |sin(}b\mathrm{)}| > 100 ~\mathrm{cm}^{-3} ~\mathrm{pc}
% \end{split}
% \end{equation}

\begin{subequations}\label{eq:thresholds}
\begin{align}
	& \mathrm{S/N} > 10  & \\
	& \mathrm{W} \leq 2^8 \times 64 \mathrm~\upmu \mathrm{s} = 16.3 ~\mathrm{ms} & \\
	& \mathrm{DM}/\mathrm{DM}_\mathrm{Galaxy} > 0.9 &
\end{align}
\end{subequations}

\noindent where the width W corresponds to $2^{\mathrm{M}}$, where M is the trial $\in (0, 1, 2, \ldots 8)$. The DM threshold was intentionally set to include high-DM candidates from sources within the Galaxy to ensure sensitivity near the theoretical DM$_\mathrm{Galaxy}$ boundary along these lines of sight. DM$_\mathrm{Galaxy}$ used in this analysis is obtained from NE2001, a model of the Galactic electron density \citep{Cordes02}. 

All pointings were searched out to a DM$_{\mathrm{max}}$ of 5000 cm$^{-3}$ pc. Extragalactic sources will have a value of DM/DM$_\mathrm{Galaxy}$ $\gtrsim$ 1 and pulsars will have DM/DM$_\mathrm{Galaxy}$ $\lesssim$ 1, if the estimated DM$_\mathrm{Galaxy}$ from NE2001 is correct. For pointings in the intermediate latitude survey the ratio of DM$_\mathrm{max}$ to DM$_\mathrm{Galaxy}$ ranges from 1.1 close to the Galactic centre to $>$ 50 at higher latitudes; we were therefore sensitive to extragalactic sources along every line of sight observed during the survey. The spatial volume probed by this search is at least the same as that in the \citet{Thornton13} analysis if not greater, depending on the maximum redshift to which FRBs are detectable.

%\section{Results}\label{sec:results}

A total of 52 candidates were identified in the 1,157 hours of the intermediate latitude survey after applying these criteria. Of these, 29 were found to be zero-DM RFI, and 23 events were caused by narrow-band RFI. No new, highly-dispersed pulses were detected. Our pipeline was also run on a fraction of the HTRU high latitude beams to search for FRBs. All four previously published FRBs were recovered in this analysis. Had thesef events occured in the intermediate latitude dataset, they would have been detected. 

\section{Discussion}\label{sec:discussion}
Although the HTRU intermediate latitude survey observed for almost twice as long as the survey reported by \citet{Thornton13}, no bursts were detected. The likelihood of finding N FRBs in our survey based on M FRBs detected at high latitudes, marginalized over the unknown FRB rate, is given by
\begin{equation}\label{eq:bayes}
P(\mathrm{N}|\mathrm{M}) = \int_0^\infty P(\mathrm{N}|\alpha \eta)P(\eta|\mathrm{M})\mathrm{d}\eta
\end{equation}
\noindent where $\eta$ is the expected number of detections, and $\alpha$ has a value of 1.88, the ratio of on-sky time of this survey compared to the Thornton survey. Here $P(\mathrm{N}|\alpha \eta)$, the distribution of the number of events given an expected number of detections, is Poissonian. Using Bayes' Theorem and a flat prior for $\eta$, the distribution for $P(\eta|\mathrm{M})$ is also Poissonian. Equation~\ref{eq:bayes} thus reduces to: 
\begin{equation}\label{eq:prob}
P(\mathrm{N}|\mathrm{M}) = \alpha^{\mathrm{N}} (1+\alpha)^{-(1+\mathrm{M} + \mathrm{N})} \frac{(\mathrm{M}+\mathrm{N})!}{\mathrm{M}! \: \mathrm{N}!} .
\end{equation}
\noindent Based on a detection of 4 FRBs in 615 hours of high latitude data the probability of detecting 0 FRBs in the intermediate latitude survey is 0.5\%, thus excluding the hypothesis that FRBs are uniformly distributed on the sky with 99.5\% confidence. 

The low probability of this result poses a critical question regarding the spatial distribution of these events. No systematic bias was introduced by the pointing position of the telescope, as both survey components covered similar distribution of telescope azimuth and elevation. The primary differences between the two HTRU survey components are Galactic effects introduced along sightlines at lower Galactic latitudes. 

Here we discuss four potential contributors to a decreased sensitivity to FRBs at $|b|<15^{\circ}$ - dispersion in the interstellar medium (ISM), interstellar scattering, sky temperature, and scintillation effects. 

\subsection{Dispersion in the ISM}\label{sec:DM}
A broadband radio pulse traveling through the ISM experiences a dispersive delay proportional to the squared wavelength of the radiation and the magnitude of the delay is a measure of the electron column density along the line of sight, DM. All radio pulsars in the Galaxy have a measured DM that can be related to distance from Earth using Galactic electron density models such as NE2001, used in most pulsar DM/distance estimates. The NE2001 model has been calibrated against nearby pulsars for which distances are known.

Using NE2001 as a model of the ionized Galaxy, we can estimate the maximum DM contribution $\mathrm{DM}_\mathrm{Galaxy}$ from the Milky Way along any line of sight. At high Galactic latitudes sightlines probe the diffuse halo of the Galaxy and $\mathrm{DM}_\mathrm{Galaxy}$ is typically less than 50 cm$^{-3}$ pc. Within the region of the intermediate latitude survey, however, Galactic dispersion contributes an average of 380 cm$^{-3}$ pc and has been measured to contribute as much as 1778 cm$^{-3}$ pc near the Galactic center \cite{GCmagnetar}. 

After subtracting DM$_\mathrm{Galaxy}$ as predicted by NE2001, the high-latitude FRBs have excess DMs between 521 and 1072, which can be attributed to the IGM and any putative host galaxy. For an FRB pulse with a DM$_{0}$ = DM$_\mathrm{host}$ + DM$_\mathrm{IGM}$ entering the Galaxy along a sightline through the intermediate latitude region the total DM observed would, on average, be of order $\sim$1500 cm$^{-3}$ pc for an FRB with DM$_0$ similar to the maximum in the known sample. We would recover this pulse as it is still below the maximum DM trial in our search (5000 cm$^{-3}$ pc) in the absence of other pulse smearing effects. 

\subsection{Scattering in the ISM}\label{sec:scattering}

Three possible scattering regimes can be considered for FRBs at cosmological distances, scattering due to the host galaxy, the IGM, and the ISM of the Milky Way. Previous studies concerned with the detectability of FRBs have assumed two extreme cases for the IGM: strong, ISM-like scattering, and no scattering \citep{Hassal13,Trott13,Lorimer13}. If IGM scattering was as strong as that of the ISM, FRB pulses would result in reduced sensitivity with current telescopes, leading to the conclusion that IGM scattering is likely weak or even un-observable \citep{JP2013}. Only one FRB in the HTRU sample, FRB110220, showed measurable scattering. 

Here we consider only the effects of Galactic scattering due to multipath propagation in the ISM, as IGM and host contributions appear to be minimal, and should have similar values for all FRBs irrespective of position. 

We scale the effects of Galactic scattering in the intermediate latitude survey region using the NE2001 model and the relationship between DM and scattering timescale from \citet{Bhat04}. This relation calculates the expected total of pulse broadening effects along any line of sight for an input DM. We use the Galactic DM$_\mathrm{Galaxy}$ obtained from NE2001 for our analysis in Section~\ref{sec:DM} as input to estimate a scattering timescale $\tau_{d}$. %For example, only 15\% of the survey pointings have pulse broadening times in excess of 10ms. 

For the majority of survey pointings ($>$85\%), we determine our measurements are still sensitive to FRB signals even in the presence of strong scattering in the ISM. 

\subsection{Sky Temperature}

A third consideration at low Galactic latitudes is the decrease in sensitivity due to sky temperature (T$_\mathrm{sky}$). At radio frequencies observations can be sensitivity limited if the standard deviation of noise fluctuations approaches the total power of the observed source \citep{PulsarHandbook}. In the analysis that follows we use the 1.4~GHz T$_\mathrm{sky}$ map from \citet{Tsky} to estimate the sky temperature for the survey region.

Excluding pointings within a few degrees of the Galactic centre, T$_\mathrm{sky}$ lies between 3 K to 30 K over the intermediate latitude survey. For $|b| > 2^{\circ}$, T$_\mathrm{sky}$ is never more than 10 K, still well below our system temperature \citep[$T_{\rm sys} = 23$\ K,][]{Keith10}. The pointings for which T$_\mathrm{sky}$ is equal to or greater than T$_\mathrm{sys}$ are already sensitivity limited due to scattering and/or the total DM contribution along the line of sight as determined in Sections~\ref{sec:DM} and~\ref{sec:scattering}.

Comparing the mean value of T$_\mathrm{sky}$ for the intermediate latitude with the mean value at high latitudes we find a difference of only 1 K between intermediate latitude survey pointings at $b > 5^{\circ}$ (4 K) and high latitude survey pointings (3 K). Sky temperature increases significantly only in the Galactic Plane where the mean temperature is closer to 10 K. In addition, T$_\mathrm{atmosphere}$ and T$_\mathrm{spillover}$ are negligible at Parkes at this frequency. Sky temperature is therefore not a significant factor in our comparison between the two survey components, and does not play a role in limiting our sensitivity, other than in regions where the survey is already limited by other Galactic factors.

\subsection{Scintillation}

The dearth of low latitude FRB events compared with high latitudes might also be explained by scintillation at high Galactic latitudes where FRB pulses may be amplified by single wideband scintles. 

For the four known FRBs, the scintillation bandwidths $\Delta\nu_{d}$ predicted using NE2001 for each line of sight are 4.8, 2.5, 5.8, and 6.1~MHz for FRB110220, FRB110626\footnote{The published name for this FRB in \citet{Thornton13} (FRB110627) is incorrect based on the UTC naming convention.}, FRB110703, and FRB120127, respectively. All are around two orders of magnitude too small to produce significant amplification. However, these values are extrapolations from the \citet{Bhat04} model fit, which has a high variance at the extremes, and may not be exact. 

Amplifications of pulsar fluxes by orders of magnitude have been observed in the 47 Tucanae and SMC pulsars \cite{Camilo}. If some FRBs are similarly amplified it may be that a fraction of those observed would not have been detected unless favourable scintillation conditions existed at the time of their arrival. In the galactic plane we know that pulsar fluxes are relatively stable \cite{Stinebring} and not amplified by diffractive scintillation to the same degree. So diffractive scintillation may help explain the larger number of FRBs we detect at high latitudes.

\subsection{Sensitivity Map}

The factors introduced in the previous sections can ultimately be combined to produce a map to determine the fraction of survey pointings in which the combination of these effects dramatically limits sensitivity. We created a simple mask to simulate the expected Galactic effect on a pulse traveling through the ISM. The effects of the temporal smearing of a pulse across the band due to dispersion, pulse broadening due to scattering, and sky temperature will combine to decrease the signal-to-noise detected for an FRB pulse from an `intrinsic' value S/N to an effective S/N$'$ as

\begin{equation}\label{eq:sensitivity}
\begin{split}
& \tau' = \sqrt{\tau_{d}^{2} +\mathrm{t}_\mathrm{disp}^{2}} \\
& \mathrm{W}' = \sqrt{\tau'^{2} + \mathrm{W}_\mathrm{int}^{2}} \\
& \mathrm{S/N}' = \mathrm{S/N} \times \frac{\sqrt{\mathrm{W}_\mathrm{int}/\mathrm{W}'}}{(1+\mathrm{T}_\mathrm{sky}/23 \mathrm{K})}
\end{split}
\end{equation}

\noindent for a receiver with T$_\mathrm{sys}$ of 23~K. Here $\tau_{d}$ is the pulse broadening time, and $\mathrm{t}_\mathrm{disp}$ is the pulse smearing time due to dispersion, which combine to give an effective scattering timescale $\tau'$, and W$_\mathrm{int}$ and W$'$ are the intrinsic and effective widths, respectively. This set of equations allows us to estimate the detectability of an FRB pulse within the parameters of our survey.

\begin{figure}
\includegraphics[width=9cm]{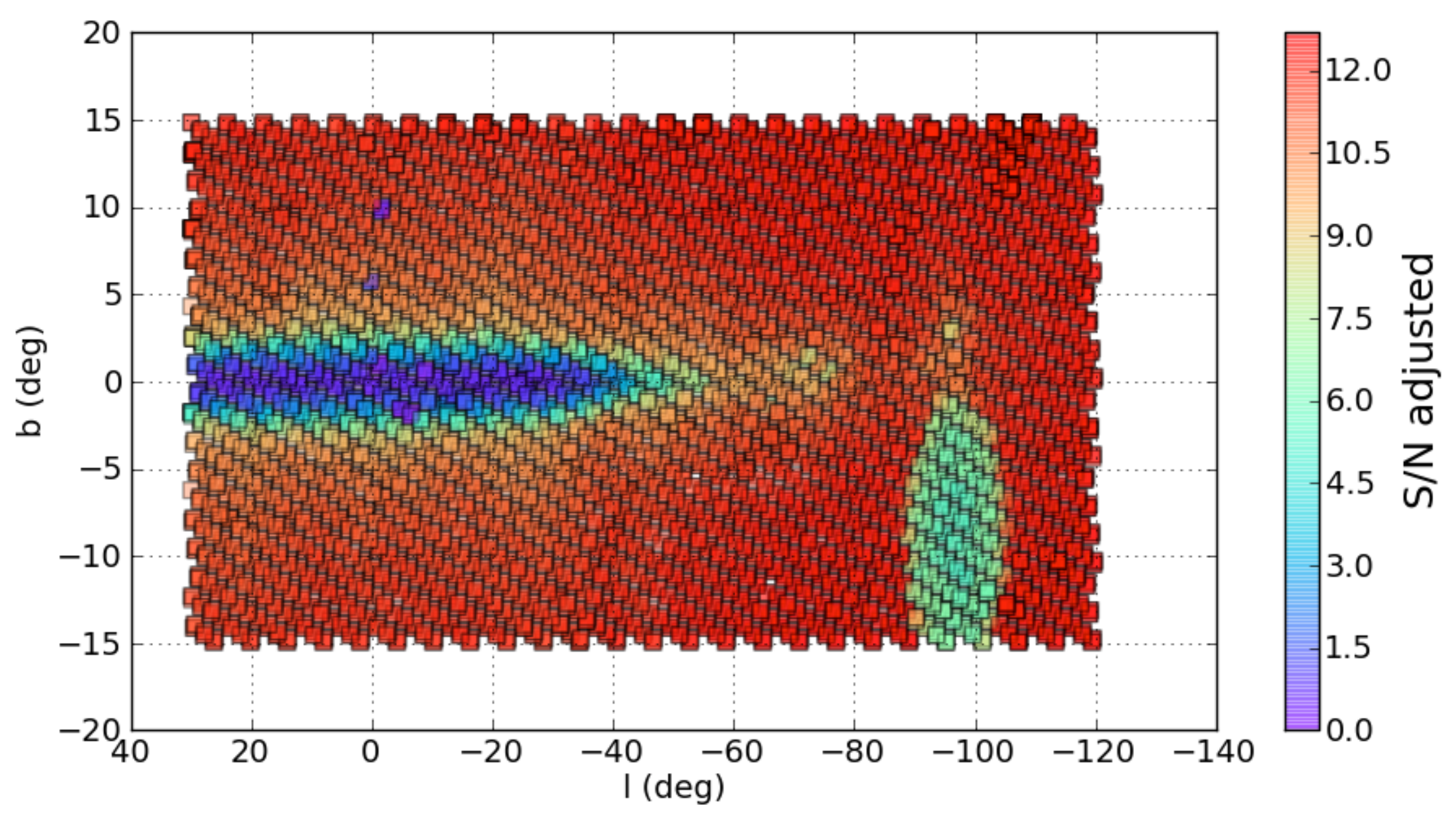}
\caption{The effective signal-to-noise, S/N$'$, of an FRB observed along all sightlines of the intermediate latitude survey. An FRB event is simulated with S/N = 13, and width = 2-ms before entering the Galaxy and the effects of dispersive smearing, interstellar scattering, and T$_\mathrm{sky}$ are taken into account to estimate the effective signal-to-noise with which the pulse would be detected with the Parkes multibeam receiver.}\label{fig:mask}
\end{figure}

From the values in Equations~\ref{eq:sensitivity} we can create a map of the intermediate latitude survey and see where a simulated FRB pulse with properties similar to the FRBs in \citet{Thornton13} falls below our signal-to-noise threshold of S/N = 10. In Figure~\ref{fig:mask} we simulate an FRB detection with S/N = 13, and a pulse width before traveling through the Galaxy, W$_\mathrm{int}$ = 2~ms. The pulse falls below the detection threshold in 20\% of all intermediate latitude pointings, primarily in high-DM regions where the dispersion smearing time and pulse broadening time are large. As expected the regions where we are least sensitive to FRBs are in the vicinity of the Galactic center, in the Galactic Plane, and through the Gum Nebula (at $(\ell, b) \sim (-100^{\circ},-10^{\circ})$).

The total time on sky of pointings not sensitive to FRBs by this metric is 231 hours of observing. This reduces the value of $\alpha$ in Equation~\ref{eq:prob} from 1.88 to  1.51  and increases the probability to 1.0\%.

\subsection{Summary}

\begin{figure}
\includegraphics[width=9cm]{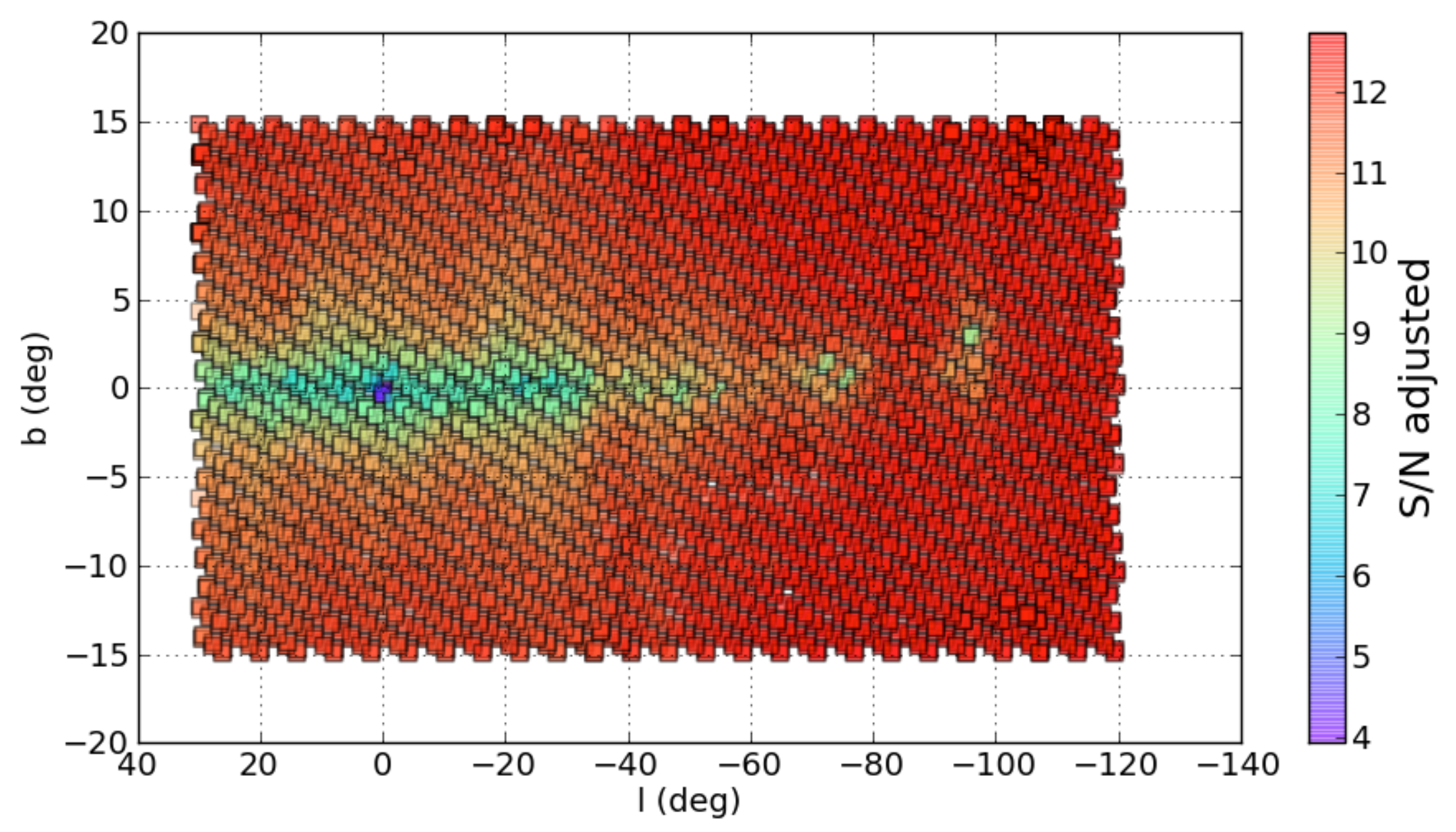}
\caption{The effective signal-to-noise, S/N$'$ of the simulated FRB from Figure~\ref{fig:mask} with no ISM scattering effects applied. Pulse smearing due to dispersion in the ISM is still accounted for. In this case the simulated FRB would drop below our detection threshold in only 14\% of survey pointings \label{fig:noScattering}}
\end{figure}

Even when taking into account Galactic effects, the probabilty of non-detection at intermediate latitudes given the current rate estimate is extremely low. We note that the \citet{Bhat04} model describes the scattering that occurs in our Galaxy; however, owing to the `lever arm' effect \cite{Hassal13}, the contribution of Galactic scattering to extragalactic sources may be significantly reduced. In the `no scattering case', the percentage of pointings no longer sensitive FRB pulses decreases to 14\% (Figure~\ref{fig:noScattering}) making our results and original predictions still more discrepant. 

Currently, the rate calculation is based on a handful of sources in only a fraction of the HTRU high latitude data. Further analysis is expected to reveal a substantial population of FRBs that will allow for a more precise rate calculation. An increased sample of FRBs is essential to resolving whether or not there is a true lack of FRBs detected through, or in the direction of, the Galactic plane.

\section{Conclusions}\label{sec:conclusions}
We conducted a single pulse search of 1,157 hours of the High Time Resolution Universe intermediate latitude survey of the Southern radio sky at 1.4~GHz with 64-$\upmu$s time resolution. We searched the data using the new GPU-based single pulse processing tool \textsc{Heimdall} over a range of incoherent DM trials from 0 to 5000 cm$^{-3}$ pc for a range of pulse widths. We searched for pulses with FRB-like measureable pulse properties with a ratio DM/DM$_\mathrm{Galaxy}$ $>$ 0.9. 

We did not detect any FRBs in the intermediate latitude survey. We verified the search pipeline on the high latitude FRBs in a blind search and all were recovered. 

Based on the detections in 615 hours of observations by \citet{Thornton13} the probability of a non-detection in the HTRU intermediate latitude survey was 0.50\%; this low probability lead us to investigate possible causes of the discrepancy. The combined contribution along sightlines through the Galactic disk of dispersion smearing, interstellar scattering, and sky temperature account for a $\sim$20\% decrease in FRB-sensitive pointings for the survey, however this is still not enough to explain the null result and only increases the probability of a non-detection to 1.0\%, thus excluding the hypothesis that FRBs are uniformly distributed on the sky with 99\% confidence. We conclude that the low probability of agreement between results at high and intermediate latitudes reveals a disagreement between the rates calculated from these two surveys. A further analysis of the HTRU high latitude data will provide a more stringent limit on all-sky FRB rates.

%the current published rate for FRB events is inconsistent with a null result at the 90\% confidence interval and the current rate estimate is too high at this confidence level.

We note that any Galactic model of FRB events must explain not only the dispersion and scattering seen in FRB pulses, but also the latitude distribution found in this work.

\section*{Acknowledgements} 
We would like to thank the referee for his or her useful comments and suggestions for this manuscript. We would also like to thank K. Lee and C. Blake for insightful discussions about statistical calculations.

The Parkes radio telescope is part of the Australia Telescope National Facility which is funded by the Commonwealth of Australia for the operation as a National Facility managed by CSIRO.

Parts of this research were conducted by the Australian Research Council Centre of Excellence for All-sky Astrophysics (CAASTRO), through project number CE110001020.


\begin{thebibliography}{}

\bibitem[\protect\citeauthoryear{{Bagchi}, Nieves \& McLaughlin}{{Bagchi} 
  et~al.}{2012}]{Bagchi12}
{Bagchi} M., {Nieves}, A.~C., {McLaughlin}, M., 2012, \mnras, 425, 2501

%\bibitem[\protect\citeauthoryear{{Barr} et~al.}{{Barr}
%  et~al.}{2013}]{hitrunNorth}
%{Barr} E.~D. et~al., 2013, \mnras, 435, 2234

\bibitem[\protect\citeauthoryear{{Bhat} et~al.}{{Bhat} et~al.}{2004}]{Bhat04}
{Bhat} N.~D.~R., {Cordes} J.~M., {Camilo} F., {Nice} D.~J.,  {Lorimer} D.~R.,
  2004, ApJ, 605, 759

\bibitem[\protect\citeauthoryear{{Burke-Spolaor} et~al.}{{Burke-Spolaor}
  et~al.}{2011a}]{Burke11}
{Burke-Spolaor} S., {Bailes} M., {Ekers} R., {Macquart} J.-P.,  {Crawford} F.,
  III, 2011a, ApJ, 727, 18

\bibitem[\protect\citeauthoryear{{Burke-Spolaor} et~al.}{{Burke-Spolaor}
  et~al.}{2011b}]{BurkeSpolaorSP}
{Burke-Spolaor} S. et~al., 2011b, MNRAS, 416, 2465

\bibitem[\protect\citeauthoryear{{Camilo} et~al.}{{Camilo}
 et~al.}{2000}]{Camilo}
 {Camilo} F., {Lorimer} D.~R., {Freire} P., {Lyne} A.~G., {Manchester} R.~N., 2000, \apj, 535, 2

\bibitem[\protect\citeauthoryear{{Cordes} \& {Lazio}}{{Cordes} \&
  {Lazio}}{2002}]{Cordes02}
{Cordes} J.~M.,  {Lazio} T.~J.~W., 2002, ArXiv Astrophysics e-prints
  (arXiv:0207156)

\bibitem[\protect\citeauthoryear{{de Oliveira-Costa} et~al.}{{de
  Oliveira-Costa} et~al.}{2008}]{Tsky}
{de Oliveira-Costa} A., {Tegmark} M., {Gaensler} B.~M., {Jonas} J., {Landecker}
  T.~L.,  {Reich} P., 2008, \mnras, 388, 247

\bibitem[\protect\citeauthoryear{{Dennison}}{{Dennison}}{2014}]{Dennison14}
{Dennison} B., 2014, ArXiv Astrophysics e-prints (arXiv:1403.2263)

\bibitem[\protect\citeauthoryear{{Eatough} et~al.}{{Eatough}
  et~al.}{2013}]{GCmagnetar}
{Eatough} R.~P. et~al., 2013, \nat, 501, 391

\bibitem[\protect\citeauthoryear{{Falcke} \& {Rezzolla}}{{Falcke} \&
  {Rezzolla}}{2013}]{Falcke}
{Falcke} H.,  {Rezzolla} L., 2014, \aap, 562, A137

\bibitem[\protect\citeauthoryear{{Hassall}, {Keane}, \& {Fender}}{{Hassall}
  et~al.}{2013}]{Hassal13}
{Hassall} T.~E., {Keane} E.~F.,  {Fender} R.~P., 2013, MNRAS

\bibitem[\protect\citeauthoryear{{Katz}}{Katz}{2014}]{KatzPerytons}
{Katz} J.~I., 2014, ArXiv Astrophysics e-prints (arXiv:1403.0637)

\bibitem[\protect\citeauthoryear{{Keane} et~al.}{{Keane}
  et~al.}{2012}]{Keane12}
{Keane} E.~F., {Stappers} B.~W., {Kramer} M.,  {Lyne} A.~G., 2012, MNRAS, 425,
  L71

\bibitem[\protect\citeauthoryear{{Keith} et~al.}{{Keith}
  et~al.}{2010}]{Keith10}
{Keith} M.~J. et~al., 2010, MNRAS, 409, 619

\bibitem[\protect\citeauthoryear{{Kocz} et~al.}{{Kocz} et~al.}{2012}]{Kocz12}
{Kocz} J., {Bailes} M., {Barnes} D., {Burke-Spolaor} S.,  {Levin} L., 2012,
  MNRAS, 420, 271

\bibitem[\protect\citeauthoryear{{Kulkarni} et~al.}{{Kulkarni} et~al.}{2014}]{Shri}
{Kulkarni} S.~R., {Ofek} E.~O., {Neill} J.~D., {Zheng} Z., {Juric} M., 2014, ArXiv Astrophysics e-prints (arXiv:1402.4766)

\bibitem[\protect\citeauthoryear{{Loeb}, {Shvartzvald}, \& {Maoz}}{{Loeb}
  et~al.}{2014}]{Loeb2014}
{Loeb} A., {Shvartzvald} Y.,  {Maoz} D., 2014, \mnras

\bibitem[\protect\citeauthoryear{{Lorimer} et~al.}{{Lorimer}
  et~al.}{2007}]{Lorimer07}
{Lorimer} D.~R., {Bailes} M., {McLaughlin} M.~A., {Narkevic} D.~J.,  {Crawford}
  F., 2007, Science, 318, 777

\bibitem[\protect\citeauthoryear{{Lorimer} et~al.}{{Lorimer}
  et~al.}{2013}]{Lorimer13}
{Lorimer} D.~R., {Karastergiou} A., {McLaughlin} M.~A.,  {Johnston} S., 2013,
  MNRAS

\bibitem[\protect\citeauthoryear{{Lorimer} \& {Kramer}}{{Lorimer} \&
  {Kramer}}{2004}]{PulsarHandbook}
{Lorimer} D.~R.,  {Kramer} M., 2004, {Handbook of Pulsar Astronomy}, Vol.~4.
\newblock Cambridge University Press

\bibitem[\protect\citeauthoryear{{Luan}}{{Luan}}{2014}]{Luan}
{Luan} J., 2014, ArXiv Astrophysics e-prints (arXiv:1401.1795)

\bibitem[\protect\citeauthoryear{{Macquart} \& {Koay}}{{Macquart} \&
  {Koay}}{2013}]{JP2013}
{Macquart} J.-P.,  {Koay} J.~Y., 2013, \apj, 776, 125

\bibitem[\protect\citeauthoryear{{Manchester} et~al.}{{Manchester}
  et~al.}{2001}]{PMPS}
{Manchester} R.~N. et~al., 2001, MNRAS, 328, 17

\bibitem[\protect\citeauthoryear{{Spitler} et~al.}{{Spitler}
  et~al.}{2014}]{Spitler14}
{Spitler} L.~G. et~al., 2014, ArXiv Astrophysics e-prints (arXiv:1404.2934)

\bibitem[\protect\citeauthoryear{{Staveley-Smith} et~al.}{{Staveley-Smith}
  et~al.}{1996}]{multibeam}
{Staveley-Smith} L. et~al., 1996, PASA, 13, 243

\bibitem[\protect\citeauthoryear{{Stinebring} et~al.}{{Stinebring} et~al.}{2000}]{Stinebring}
 {Stinebring} D. et~al., 2000, \apj, 539, 1

\bibitem[\protect\citeauthoryear{{Thornton} et~al.}{{Thornton}
  et~al.}{2013}]{Thornton13}
{Thornton} D. et~al., 2013, Science, 341, 53

\bibitem[\protect\citeauthoryear{{Trott}, {Tingay}, \& {Wayth}}{{Trott}
  et~al.}{2013}]{Trott13}
{Trott} C.~M., {Tingay} S.~J.,  {Wayth} R.~B., 2013, ApJ Letters, 776, L16

\end{thebibliography}
\end{document}